\documentclass{pic2012}
\usepackage{ifpdf}
\ifpdf
\usepackage{wrapfig}
\usepackage{float}
\usepackage[rflt]{floatflt}
\usepackage{epsfig}
\usepackage{cite}

\def\TCop{\textsuperscript{\textcopyright}}

\def\runauthor{S. Alekhin, J. Bl\"umlein, S. Moch}
\def\shorttitle{Parton Distributions and $\alpha_s$ for the LHC}
\begin{document}

\title{
{\footnotesize \rm DESY 13-021, DO-TH 13/05,  SFB/CPP-13-012, LPN13-016 \hfill {\tt arXiv:1303.1073 
[hep-ph]}}
\\
Parton Distributions and $\alpha_s$ for the LHC}

\author{S. Alekhin$^{\rm a,b}$, J. Bl\"umlein$^{\rm a}$\footnote{Speaker; 
Johannes.Bluemlein@desy.de}, S. Moch$^{\rm a,c}$}

\address{$^{\rm A}$DESY, Platanenallee 6, D--15738 Zeuthen, Germany\\
         $^{\rm B}$Institute for High Energy Physics,142281 Protvino, Moscow region, 
         Russia\\
         $^{\rm C}$II. Institut f\"ur Theoretische Physik, Universit\"at Hamburg
         Luruper Chaussee 149, D-22761 Hamburg, Germany}

\maketitle

\abstracts{
We report on recent determinations of NNLO parton distributions and of 
$\alpha_s(M_Z)$ based on the world deep-inelastic data, supplemented by 
collider data. Some applications are discussed for semi-inclusive 
processes at the LHC.}

\section{Introduction} 

\noindent
The precise knowledge of the parton distribution functions (PDFs) and the strong coupling constant
$\alpha_s(M_Z^2)$ is of central importance for the description of any high energy
reaction, in which hadrons are involved. In this way both the search for new particles,
including the Higgs boson(s), and the precision description of the associated scattering
cross sections at the LHC do crucially depend on these quantities. This also applies
for detailed investigations of other heavy systems, which can now be explored in greater
detail, as $W^\pm, Z^0$-, jet- and top-quark production. On the other hand, both the world precision 
data on deep-inelastic scattering (DIS) and precise enough hadron collider data allow to fit the 
parton 
distribution functions and $\alpha_s(M_Z^2)$ in a correlated way and to test QCD. The 
corresponding analyses have to be based on the most accurate theoretical descriptions
available. In case of deep-inelastic scattering the massless contributions to the structure
functions are known to NNLO \cite{MVV}, while the heavy quark contributions are presently 
available to NLO \cite{HEAV2}. At the 3-loop level in the large-$Q^2$ region a series of Mellin 
moments for the massive operator matrix elements has been calculated in \cite{Bierenbaum:2009mv}. 
All logarithmic contributions for the charm quark contributions to the structure function 
$F_2(x,Q^2)$ were calculated in \cite{Bierenbaum:2010jp}. Recently, an interpolation between 
the threshold and high $Q^2$-region for the heavy flavor Wilson coefficients has been given 
in \cite{Kawamura:2012cr}, also based on the moments in Ref.~\cite{Bierenbaum:2009mv} and the 
known small-$x$ contributions. 

Currently there are results from six NNLO PDF-analyses available, four of which were published, 
with most recent results given in 
\cite{JimenezDelgado:2008hf,Martin:2009iq,Alekhin:2012ig,Ball:2011uy,HERAPDF,CTEQ12}.
Due to the strong correlations in the fit the non-perturbative parameters of the PDFs and 
$\alpha_s(M_Z^2)$ have to be determined together. A detailed account of the systematic 
errors of the different experiments is necessary. In part of the phase space target mass effects 
\cite{Georgi:1976ve} and higher twist terms have to be accounted for. For a recent survey on the 
theory of deep-inelastic scattering, see e.g.~\cite{Blumlein:2012bf}. Since not all data sets are of
the same quality it may be helpful to refer to those of the highest quality first and to carefully
add other data sets then, rather than fitting a wide host of data of quite different quality. 
Proceeding in this way, our focus is narrower but also sharper~\cite{GW1}.
In the following we give a brief survey on the status of the twist-2 parton distribution functions, 
on $\alpha_s(M_Z^2)$, and on hard scattering cross sections at the LHC, which start to develop 
sensitivity on the PDFs. 
\section{Parton Distribution Functions} 

\noindent
During the last years various updates on the PDFs at NNLO (and at NLO) have been given by
different groups \cite{JimenezDelgado:2008hf,Martin:2009iq,Alekhin:2012ig,Ball:2011uy,HERAPDF,CTEQ12}.
Here it is mandatory to refer to the combined HERA data \cite{Aaron:2009aa}, not yet included by all 
of the groups. In Fig.~1 we compare several PDFs at NNLO with those being obtained
in the recent analysis \cite{Alekhin:2012ig}.
\restylefloat{figure}
\begin{figure}[H]\centering
\includegraphics[scale=0.6,angle=0]{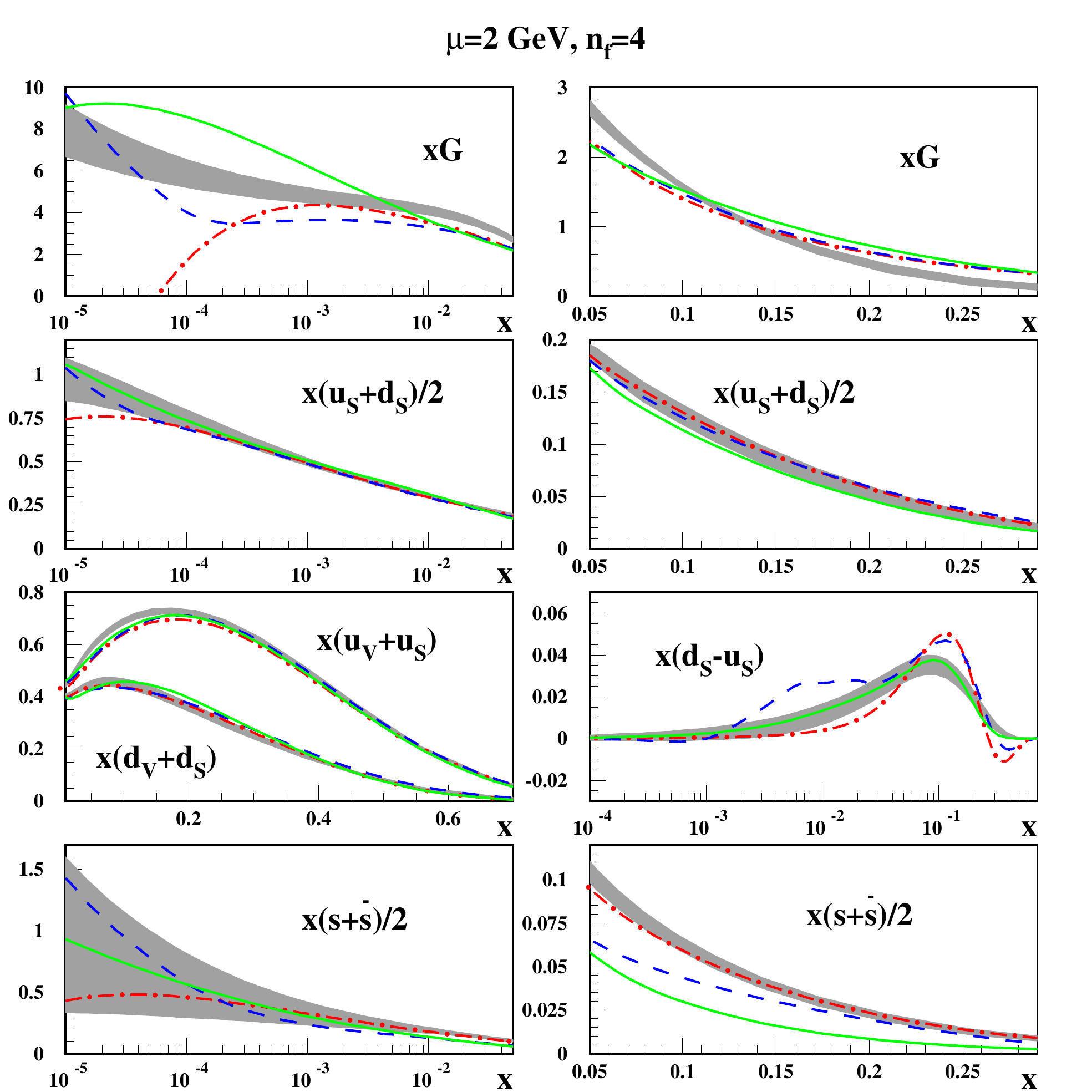}
\caption[]
{\footnotesize The $1\sigma$ band for the 4-flavor NNLO ABM11 PDFs at the scale of $\mu$ = 2 GeV 
versus $x$ 
obtained in \cite{Alekhin:2012ig} (shaded area) compared with the ones obtained by other groups 
(solid lines: JR09 \cite{JimenezDelgado:2008hf}, dashed dots: MSTW \cite{Martin:2009iq}, dashes: 
NN21 \cite{Ball:2011uy}); from Ref.~\cite{Alekhin:2012ig} \TCop (2012) by the American Physical 
Society.}
\end{figure}
\noindent
In \cite{Alekhin:2012ig} the shapes of the PDFs were parameterized by
$xf(x) = A x^{a+P(x)}(1-x)^b,~~P(x) = x(\gamma_1 + \gamma_2 x + \gamma_3 x^2)$,
resp.
$P_{\rm us}(x) = (1+\gamma_3 \ln(x))(1+ \gamma_1 x + \gamma_2 x^2)$~.
The fit parameters of the PDFs obtained are listed in Tab.~1.
\begin{figure}[t]\centering
\includegraphics[scale=0.7,angle=0]{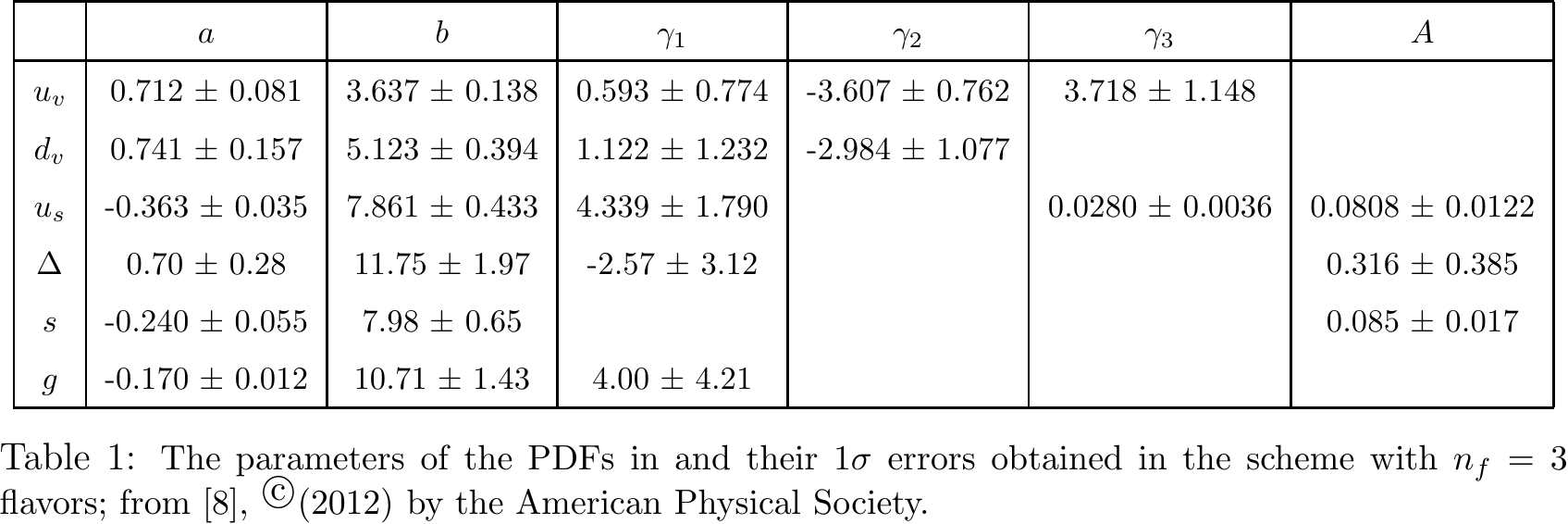}
\end{figure}
\begin{wrapfigure}{r}{0.45\textwidth}
\includegraphics[width=0.4\textwidth]{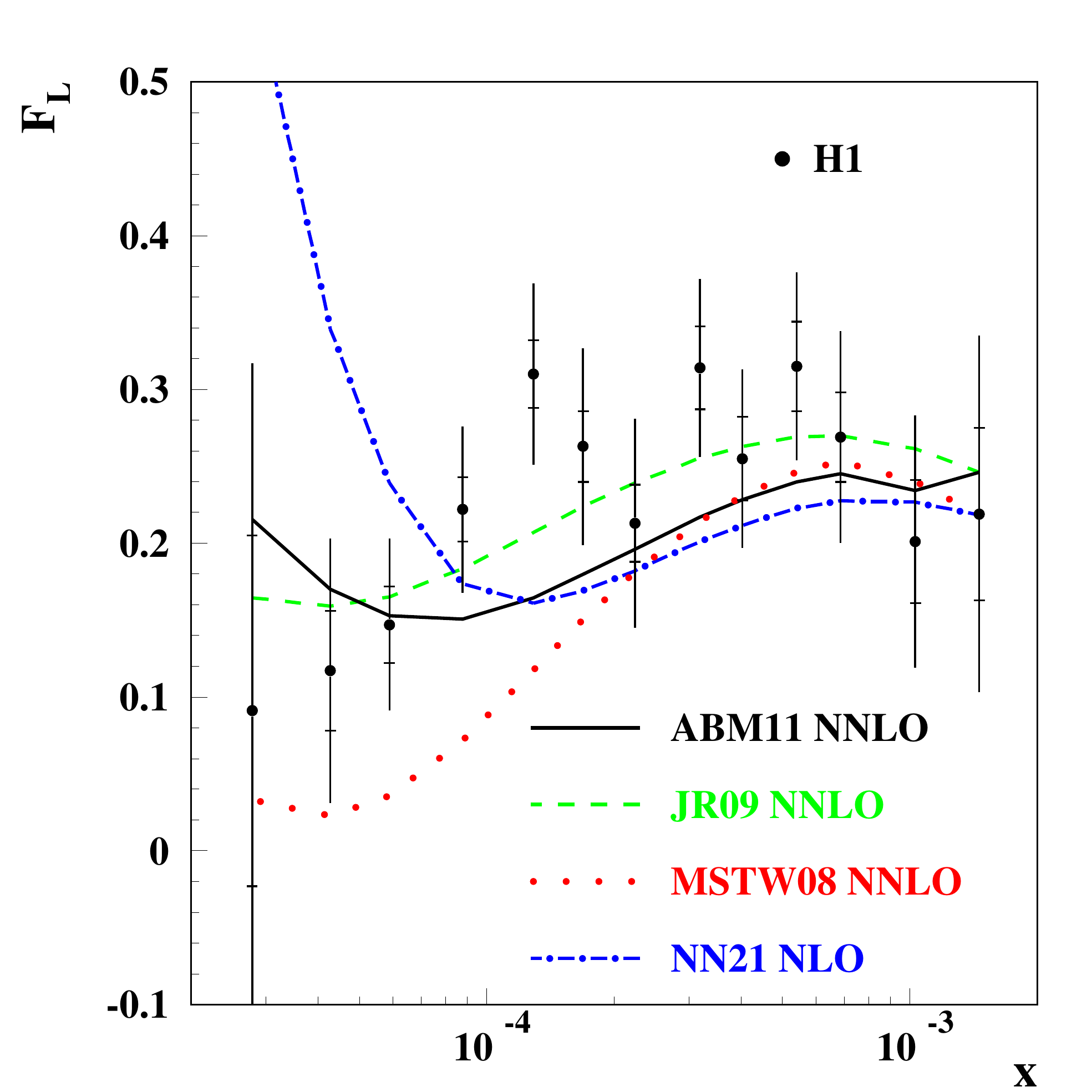}
{\footnotesize Figure~2:~The data on $F_L$ versus $x$ obtained by the H1
     collaboration~\cite{Collaboration:2010ry} 
     confronted with the 3-flavor scheme NNLO predictions based on the 
     different PDFs (solid line: ABM11 \cite{Alekhin:2012ig}, dashes: 
     JR09~\cite{JimenezDelgado:2008hf}, dots: MSTW~\cite{Martin:2009iq}). 
     The NLO predictions based on the 3-flavor NN21 
     PDFs~\cite{Ball:2011mu} are given for comparison (dashed dots). 
     The value of $Q^2$ for the data points and the curves in the plot 
     rises with $x$ in the range of $1.5$ to $45$~GeV$^2$; from Ref.~\cite{Alekhin:2012ig}, 
\TCop  (2012) by the American Physical Society. \label{fig:fl}}
\end{wrapfigure}
Also the full error correlation matrices are presented in Ref.~\cite{Alekhin:2012ig}.
While the valence and $u,d$-sea quark distributions do widely agree for the +-combinations, 
there are still significant differences in the gluon distribution at the starting scale of $\mu = 
2$~GeV between different groups. In case of MSTW \cite{Martin:2009iq} the gluon distribution takes 
negative values at 
low $x$. The measurement of $F_L(x,Q^2)$ can partly clarify the situation since it is carried out at 
lower scales of $Q^2$. In Fig.~2 predictions  from four PDF-sets are compared. MSTW~\cite{Martin:2009iq}
yields lower values in the small-$x$ region, while NN21 \cite{Ball:2011mu} gives high values. ABM11 
\cite{Alekhin:2012ig} and JR09 \cite{JimenezDelgado:2008hf} describe these data very well. To draw a 
final conclusion, more accurate data would be needed, however. They are expected to come from 
experiments 
at a future electron-ion collider \cite{Boer:2011fh}  or at LHeC \cite{AbelleiraFernandez:2012cc}.
There are some differences visible in the $\bar{d}-\bar{u}$ distributions and the strange quark
distribution. The present and upcoming LHC data on $W^\pm, Z^0$ boson production and the 
off-resonance Drell-Yan (DY) process may help to improve the situation here. The presently worst 
known 
distribution is that of the $s$-quark, while the $c$- and $b$-quark distributions (in some scheme to 
be specified) are driven by the gluon distribution mainly.

In the analysis \cite{Alekhin:2012ig} we included data down to scales of $Q^2 = 2.5$ GeV$^2$. In the
lower $W^2$-region contributions of twist-4 operators are found, which were parameterized by a
phenomenological shape $H(x)/Q^2$ additively. In Tab.~2 the corresponding fit results are given.
\restylefloat{figure}
\begin{figure}[H]\centering
\includegraphics[scale=0.7,angle=0]{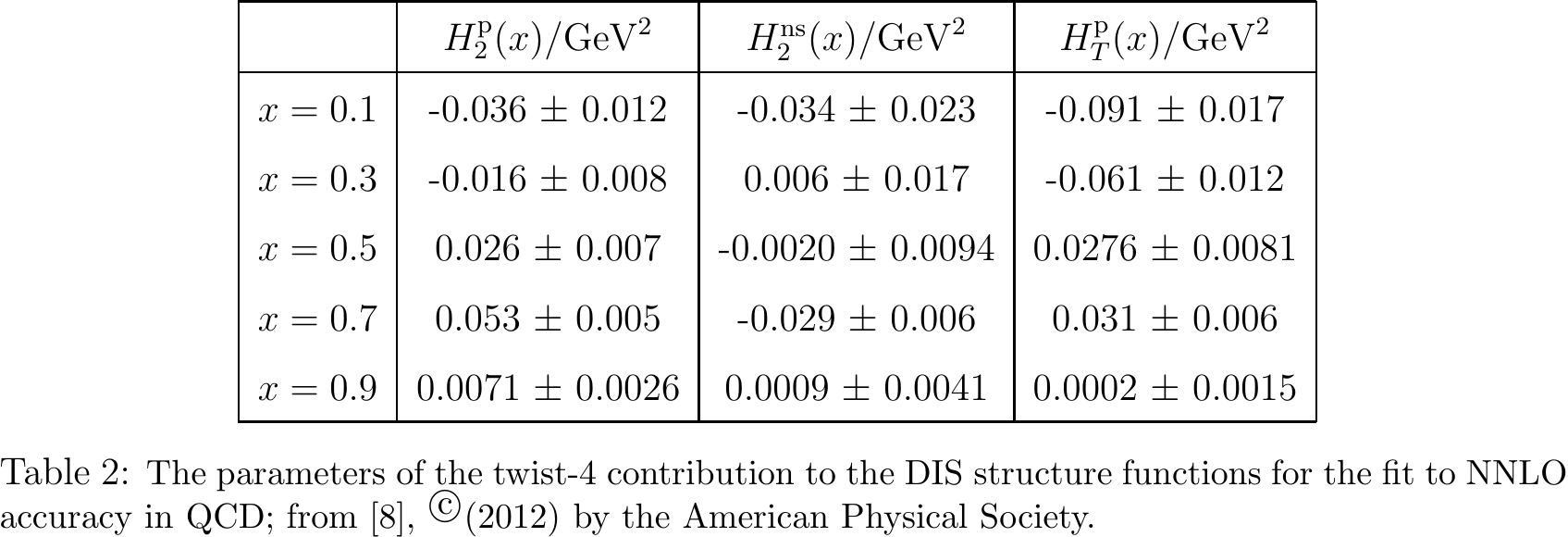}
\end{figure}
\section{The Strong Coupling Constant $\alpha_s(M_Z^2)$} 

\noindent
The strong coupling constant $\alpha_s(M_Z^2)$ is measured together with the parameters of the PDFs,
the heavy quark masses
$m_c$ and $m_b$, and the higher twist parameters within the analysis \cite{Alekhin:2012ig}. The 
present experimental accuracies of $O(1\%)$ require NNLO corrections, since at NLO the scale 
uncertainties amount to $O(5\%)$, cf.~\cite{Blumlein:1996gv}. 
In Tab.~3
we compare the response of the data sets from BCDMS \cite{BCDMS}, NMC \cite{NMC}, SLAC \cite{SLAC},
HERA \cite{Aaron:2009aa} and the Drell-Yan data \cite{DY} at NLO and NNLO and to the NLO values 
given in earlier experimental analyses. Within the experimental errors an entirely consistent 
picture
arises. Moreover, the new results agree with those of 
Refs.~\cite{Alekhin:2009ni,Blumlein:2006be,Blumlein:2012se,Gluck:2006yz,JimenezDelgado:2008hf}.
\restylefloat{figure}
\begin{figure}[H]\centering
\includegraphics[scale=0.7,angle=0]{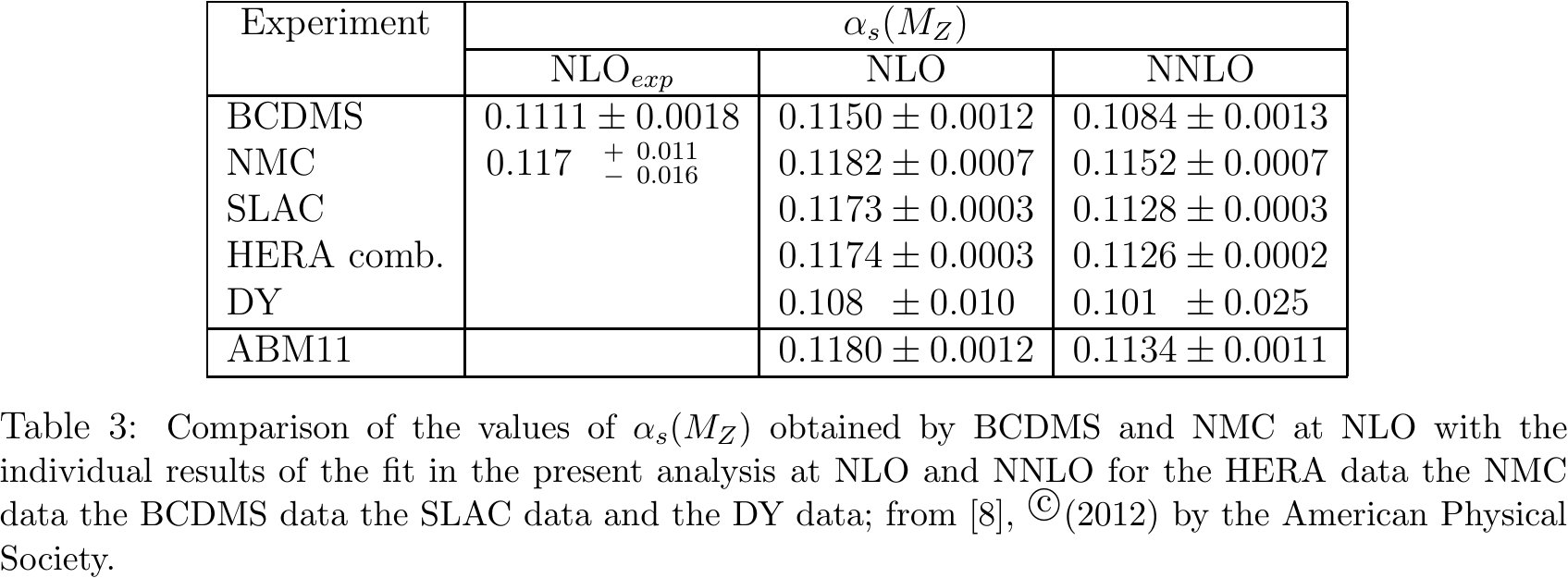}
\end{figure}
\vspace*{-6mm}
\noindent
As very well-known, both JR \cite{Gluck:2007ck} and ABM \cite{Alekhin:2011cf,Alekhin:2012ig}, along 
with other groups, carried 
out systematic fits including both jet data from the Tevatron and in \cite{Alekhin:2012ig} also from 
LHC\footnote{Contrary statements, as given in \cite{Ball:2012wy,FW}, are incorrect; 
see Ref.~\cite{Alekhin:2012ig} for all details. It is very problematic to call present fits including 
jet data and the world DIS data  NNLO analyses, since the corresponding jet scattering cross sections  
are available in NLO {\it only}. In Ref.~\cite{Alekhin:2012ig} we have accounted for 
threshold resummation contributions beyond NLO, however.}. 
{\small \begin{table}[h!]\centering
\begin{tabular}{|l|c|c|c|}
\hline
\multicolumn{1}{|c|}{Experiment} &
\multicolumn{3}{c|}{$\alpha_s(M_Z)$} \\
\cline{2-4}
\multicolumn{1}{|c|}{ } &
\multicolumn{1}{c|}{NLO$_{exp}$} &
\multicolumn{1}{c|}{NLO} &
\multicolumn{1}{c|}{NNLO$^*$} \\
\hline 
D0  1 jet             & $0.1161~^{+~0.0041}_{-~0.0048}$      
                      & $0.1190 \pm 0.0011$ & $0.1149 \pm 0.0012$ \\
D0  2 jet             &      & $0.1174 \pm 0.0009$ & $0.1145 \pm 0.0009$ \\   
CDF 1 jet (cone)      &      & $0.1181 \pm 0.0009$ & $0.1134 \pm 0.0009$ \\   
CDF 1 jet ($k_\perp$)  &      & $0.1181 \pm 0.0010$ & $0.1143 \pm 0.0009$ \\   
\hline
ABM11                 &      & $0.1180 \pm 0.0012$ & $0.1134 \pm 0.0011$ \\   
\hline 
\end{tabular} 
\renewcommand{\arraystretch}{1}  
\setcounter{table}{3}
\caption{ \small
\label{tab:alphas2}
Comparison of the values of $\alpha_s({M_Z})$ obtained by D0 
with the ones based on including individual 
data sets of Tevatron jet data 
into the analysis at NLO. 
The NNLO$^*$ fit refers to the NNLO analysis of the DIS and DY data together with 
the NLO and soft gluon resummation corrections (next-to-leading logarithmic accuracy) 
for the 1 jet inclusive data. 
}
\end{table}}

\noindent
In Tab.~4 we summarize the results 
including individual sets
of Tevatron jet data \cite{TEVjet}. Both the results in NLO and NNLO are 1$\sigma$ compatible with our 
values on
$\alpha_s(M_Z^2)$ obtained without these data. In case we include the ATLAS jet data
\cite{Aad:2011fc} into the analysis along with the DIS and Drell-Yan data we obtain $\alpha_s(M_Z^2) = 
0.1141\pm 0.0008$, demanding $p_\perp^{\rm jet} > 100$~GeV. Here the jet cross section was described 
at NLO 
adding threshold resummation, see \cite{Alekhin:2012ig} for details.

The ATLAS and CMS jet data span a wider kinematic range than those of Tevatron and will allow very 
soon even more accurate measurements. At the moment NLO QCD analyses are carried out with 
first results being obtained in \cite{Malaescu:2012ts,KR}. The values of $\alpha_s(M_Z^2)$ come out 
rather low. Including the scale uncertainties the following NLO values are obtained for the 3/2 jet
ratio at CMS
\begin{eqnarray}
\label{jet1}
\alpha_s(M_Z^2) &=& 0.1143 \pm 0.0064~~[0.1191 \pm 0.0006]~~[37] 
~~{\rm NNPDF21} \\
\alpha_s(M_Z^2) &=& 0.1130 \pm 0.0080~~[0.180~{(\rm favored~value)}]~~[38]~~{\rm CT10} \\
\alpha_s(M_Z^2) &=& 0.1135 \pm 0.0096~~[
0.1202 {\small \begin{array}{c} +0.0012 \\ 
-0.0015 \end{array}}]~~[39]~~{\rm MSTW08}~, 
\label{jet3}
\end{eqnarray} 
varying the corresponding PDF-sets given for a wider range in $\alpha_s(M_Z^2)$ \cite{KR}. Note that 
the minima
in $\chi^2$ found by the fitting groups 
are reached at much higher values of $\alpha_s(M_Z^2)$ (quoted in parenthesis). 
In the near future the scale uncertainties in 
(\ref{jet1}--\ref{jet3}) will significantly diminish upon the arrival of the NNLO corrections.
A comparable NLO value has been reported using ATLAS jet data \cite{Malaescu:2012ts}
$\alpha_s(M_Z^2) = 0.1151 \pm 0.0050~{\rm (exp.)} {\small \begin{array}{c} +0.0080 \\ 
-0.0073\end{array}}
~{\rm (th.)}$ Recently the jet energy-scale error has been improved by CMS \cite{KR}, 
leading to a significant reduction of the experimental error. Moreover, the gluonic NNLO corrections
for jet-production are now available \cite{GLOVER}, with the other NNLO terms to be expected later 
this year. The gluonic NNLO $K$-factor is positive. As shown in Fig.~2 of Ref.~\cite{GLOVER} the scale dependence for $\mu = \mu_F = \mu_R$
behaves flat over a wide range of scales. It is expected that also the error due to scale 
variation will turn out to be very small. Therefore, a definite answer on $\alpha_s(M_Z^2)$ 
at NNLO
from both ATLAS and CMS alone using jet measurement is imminent. It should be mentioned that in this 
determination both $\alpha_s(M_Z^2)$ and the gluon distribution shall be fitted due to the strong
correlations of both quantities, which also will yield an independent input on $xG(x,\mu^2)$
based on LHC data only and will allow for interesting comparisons with the gluon distributions  
determined by other means.
\restylefloat{figure}
\begin{figure}[H]\centering
\includegraphics[scale=0.7,angle=0]{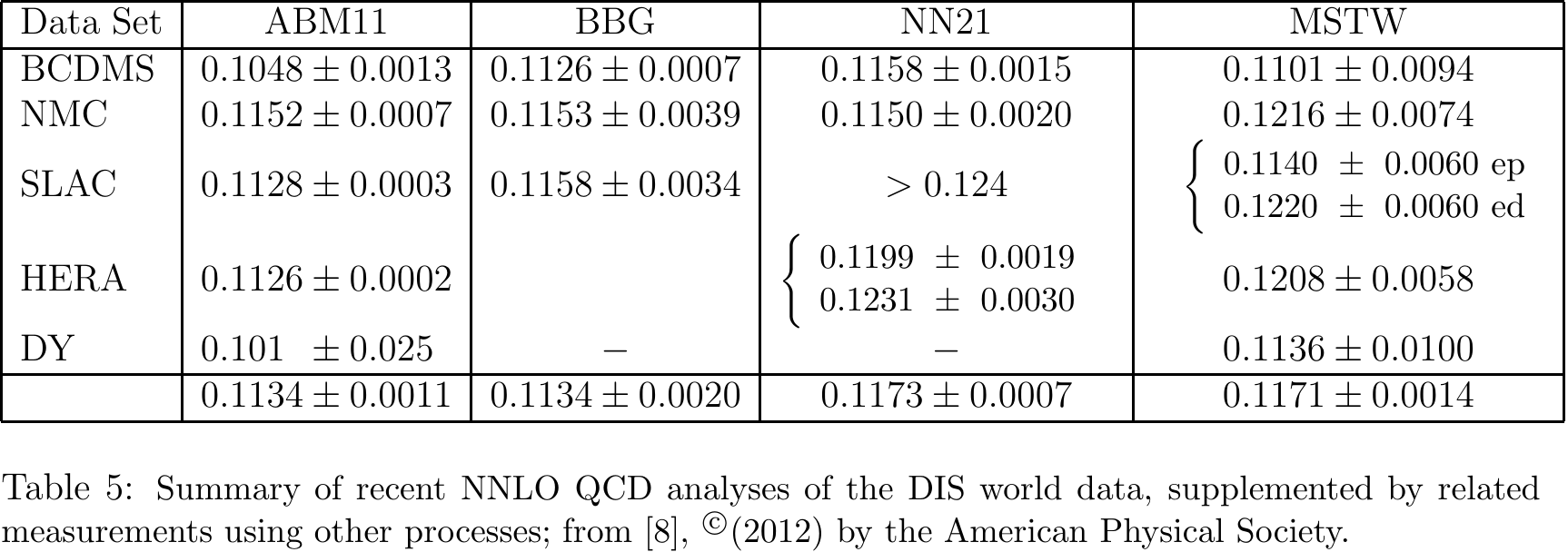}
\end{figure}
\noindent
Let us compare the results of the NNLO fits ABM11 \cite{Alekhin:2012ig}, BBG \cite{Blumlein:2006be}, NN21 
\cite{Ball:2011us} and MSTW08 \cite{Martin:2009bu} on $\alpha_s(M_Z^2)$ w.r.t. their 
individual response to the deep-inelastic and Drell-Yan data sets used, cf.~Tab.~5.
While the $\alpha_s(M_Z^2)$ values obtained in the ABM11 and BBG analyses lead to 1$\sigma$ 
consistent values between the final value of $\alpha_s(M_Z^2) = 0.1134$ and those of the individual 
data sets, for NN21 and 
MSTW08 variations are being observed. NN21 agrees with ABM11 and BBG for the 
NMC data but reports a higher value for BCDMS, with the converse for MSTW08. Both NN21 and MSTW08 come 
up with a larger value for the HERA data, however, MSTW08 does not include the combined HERA data 
yet.
For the SLAC data MSTW08 agrees with ABM11 and BBG for the $ep$ data but obtains a much larger value 
for the $ed$ data, while NN21 obtains $\alpha_s(M_Z^2) > 0.124$ in case of the SLAC data.
Because of the large errors, the Tevatron Drell-Yan data are not very decisive on the value of 
$\alpha_s(M_Z^2)$. Despite the similar final values of $\alpha_s(M_Z^2)$ obtained by NN21 and 
MSTW08, those for the individual data sets vary significantly.

We outlined in \cite{Alekhin:2011ey} that a consistent $F_L$-treatment for the NMC data and the 
BCDMS-data, cf.~\cite{Blumlein:2006be}, is necessary and will lead to a change of the value of 
$\alpha_s(M_Z^2)$. Furthermore, we analyzed the sensitivity on kinematic cuts applied to remove  
higher twist effects. In the flavor non-singlet case this can be achieved cutting for $W^2 >
12.5$~GeV$^2$, cf.~\cite{Blumlein:2006be}. As we have shown in Section~2, there are also higher twist 
contributions in the lower $x$-region. They can be removed applying a cut $Q^2 > 10$ GeV$^2$, which, 
however, is not applied by NN21 and MSTW08. We performed a fit ignoring the higher twist terms and 
allowed for the range of data down to values of $Q^2 > 2.5$~GeV$^2$,~\cite {Alekhin:2012ig}.
Here we obtain $\alpha_s(M_Z^2) = 0.1191 \pm 0.0016$, very close to the values found by NN21 and 
MSTW08. 

In Tab.~6 a general overview on the values of $\alpha_s(M_Z^2)$ at NNLO is given, with a few
determinations effectively at N$^3$LO in valence analyses \cite{Blumlein:2006be,Blumlein:2012se},
and the hadronic $Z$-decay \cite{Baikov:2012er}. The BBG, BB, GRS, ABKM, JR, and ABM11 analyses find 
lower 
values of $\alpha_s(M_Z^2)$ with errors at the 1--2\% level, while NN21 and MSTW08 find larger 
values 
with comparable accuracy to the former ones, as has been discussed before. We also add a yet 
unpublished 
NNLO value from CT10 quoting a significantly larger error. In the analysis of thrust in $e^+e^-$ data 
two groups find low values, also with errors at the 1\% level. Higher values of $\alpha_s(M_Z^2)$ 
are 
found for the $e^+e^-$ 3-jet rate, the hadronic $Z$-decay, and $\tau$-decay. $\alpha_s(M_Z^2)$ has 
also been determined in different lattice simulations to high accuracy. The N$^3$LO values for 
$\alpha_s(M_Z^2)$ in the valence analyses \cite{Blumlein:2006be,Blumlein:2012se} yield slightly larger  
values than at NNLO, but are fully consistent with these values within errors.
\begin{table}[h!]\centering
\renewcommand{\arraystretch}{1}
\begin{tabular}{|l|l|l|}
\hline
\multicolumn{1}{|c|}{ } &
\multicolumn{1}{c|}{$\alpha_s({M_Z})$} &
\multicolumn{1}{c|}{  } \\
\hline
BBG      & $0.1134~^{+~0.0019}_{-~0.0021}$
         & {\rm valence~analysis, NNLO}  [25] 
\\[0.5ex]
BB       & $0.1132  \pm 0.0022$
         & {\rm valence~analysis, NNLO}  [26] 
\\
GRS      & $0.112 $ & {\rm valence~analysis, NNLO}  [27] 
\\
ABKM           & $0.1135 \pm 0.0014$ & {\rm HQ:~FFNS~$n_f=3$} [24] 
\\
ABKM           & $0.1129 \pm 0.0014$ & {\rm HQ:~BSMN-approach} 
[24] 
\\
JR       & $0.1124 \pm 0.0020$ & {\rm
dynamical~approach} [6] 
\\
JR       & $0.1158 \pm 0.0035$ & {\rm
standard~fit}  [6] 
\\
ABM11            & $0.1134\pm 0.0011$ &  [8] 
\\
MSTW & $0.1171\pm 0.0014$ &  [39] 
\\
NN21 & $0.1173\pm 0.0007$ &  [41] 
\\
CT10 & $0.118\phantom{0} \pm 0.005$  &  [11] 
\\
\hline
Abbate et al.& {{$0.1135 \pm 0.0011 \pm 0.0006$}} & {\rm
$e^+e^-$~thrust}~[43] 
\\
Gehrmann et al.& $0.1131~^{+~0.0028}_{-~0.0022}$
& {\rm
$e^+e^-$~thrust}~[44] 
\\
\hline
3 jet rate   & $0.1175 \pm 0.0025$ & Dissertori et al. 2009 [45] 
\\
Z-decay      & $0.1189 \pm 0.0026$ & BCK 2008/12  (N$^3$LO) [46,47] 
\\
$\tau$ decay & $0.1212 \pm 0.0019$ & BCK 2008               [46] 
\\
$\tau$ decay & $0.1204 \pm 0.0016$ & Pich 2011              [48] 
\\
$\tau$ decay & $0.1191 \pm 0.0022$ & Boito et al. 2012      [49] 
\\
\hline
lattice      & $0.1205 \pm 0.0010$ & PACS-CS 2009 (2+1 fl.) [50] 
\\
lattice      & $0.1184 \pm 0.0006$ & HPQCD 2010             [51] 
\\
lattice      & $0.1200 \pm 0.0014$ & ETM 2012 (2+1+1 fl.)   [52] 
\\
lattice      & $0.1156 \pm 0.0022$ & Brambilla et al. 2012 (2+1 fl.)   [53] 
\\
\hline
BBG & $0.1141~^{+~0.0020}_{-~0.0022}$
& {\rm valence~analysis, N$^3$LO$(^*)$}  [25] 
\\[0.5ex]
BB & $0.1137 \pm 0.0022$
& {\rm valence~analysis, N$^3$LO$(^*)$}  [26] 
\\
\hline
{world average} & {$
0.1184 \pm 0.0007$  } & [54] 
(2009)
\\
                & {$
0.1183 \pm 0.0010$  } & [48] 
(2011)
\\
\hline
\end{tabular}
\renewcommand{\arraystretch}{1}   
\setcounter{table}{5}
\caption{
\small
\label{tab:alphas7}
Summary of recent NNLO QCD analyses of the DIS world data, supplemented by related measurements
using other processes; from [8], \TCop (2012) by the American Physical Society.}
\end{table}

Finally we would like to mention, that the present DIS world data together with the 
$F_2^{c\bar{c}}(x,Q^2)$ data, are competitive in the determination of the charm quark mass.
In a recent analysis \cite{Alekhin:2012vu} the value of $m_c(m_c) = 1.24 \pm 0.03~{\rm (exp.)} 
{\small 
\begin{array}{c} +0.04 \\ -0.00 \end{array}}$ (th.) is obtained at NNLO. The analysis in 
Ref.~\cite{Alekhin:2012vu} is presently the only one, in which all known higher order heavy flavor
corrections to deep-inelastic scattering have been considered. Let us note, that there is still 
a wide spectrum in the way heavy flavor corrections are accounted for in
\cite{JimenezDelgado:2008hf,Martin:2009iq,Alekhin:2012ig,Ball:2011uy,HERAPDF,CTEQ12}
w.r.t. the present knowledge on the side of theory.
\section{Hard Scattering Cross Sections at the LHC} 

\noindent
The presently obtained precision predictions on the PDFs at NNLO may be used 
to predict different scattering cross sections being measured at the LHC. This
applies in particular to those cross sections, resp. cross section ratios, which 
can be measured at high precision, cf.~\cite{Alekhin:2010dd}.
Here observables of central importance are the $W^\pm/Z^0$ cross section ratios.
In Fig.~3 we show  results obtained by ATLAS \cite{Aad:2011dm}.
\begin{figure}[b]\centering
\includegraphics[scale=0.30,angle=0]{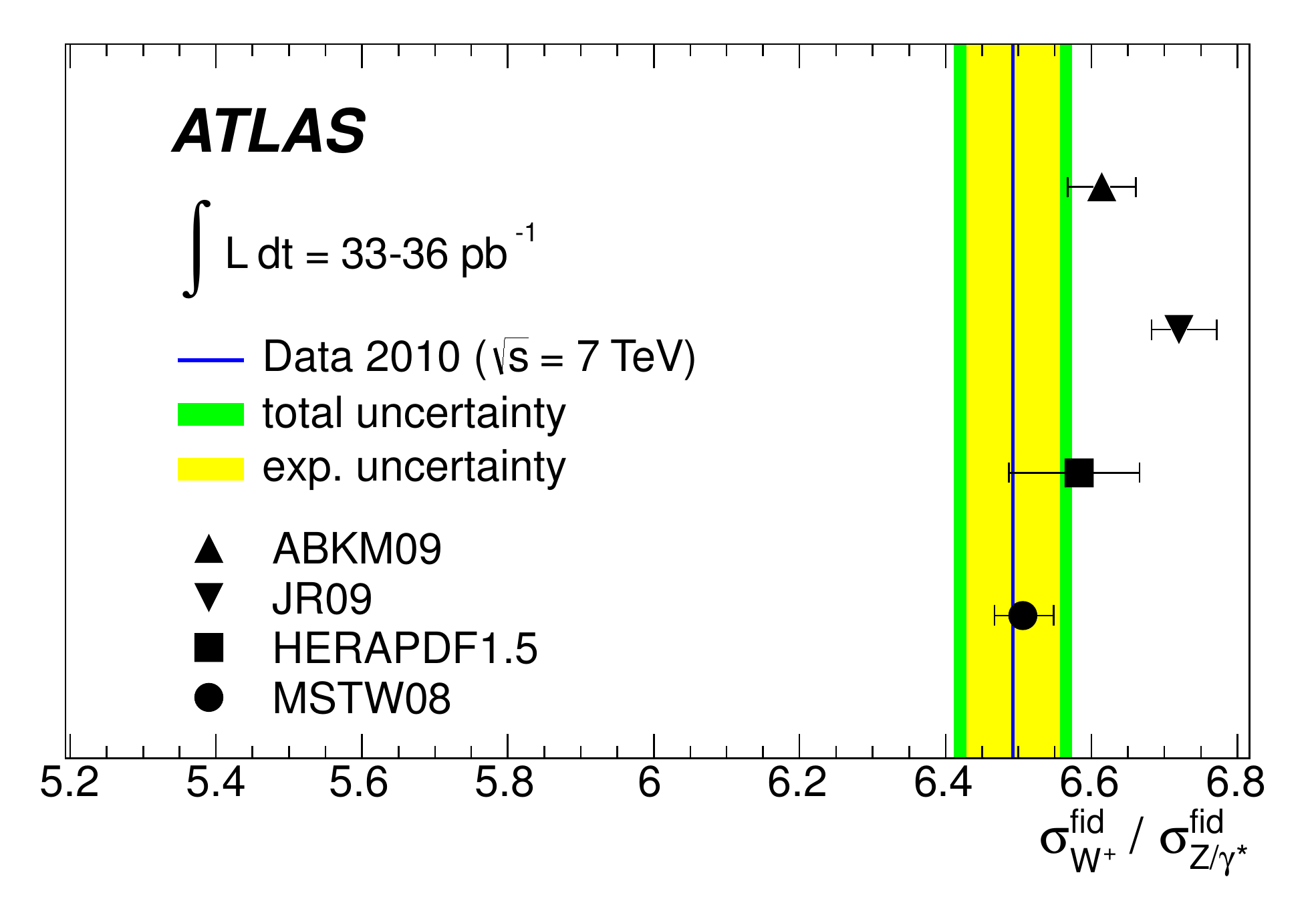}
\includegraphics[scale=0.30,angle=0]{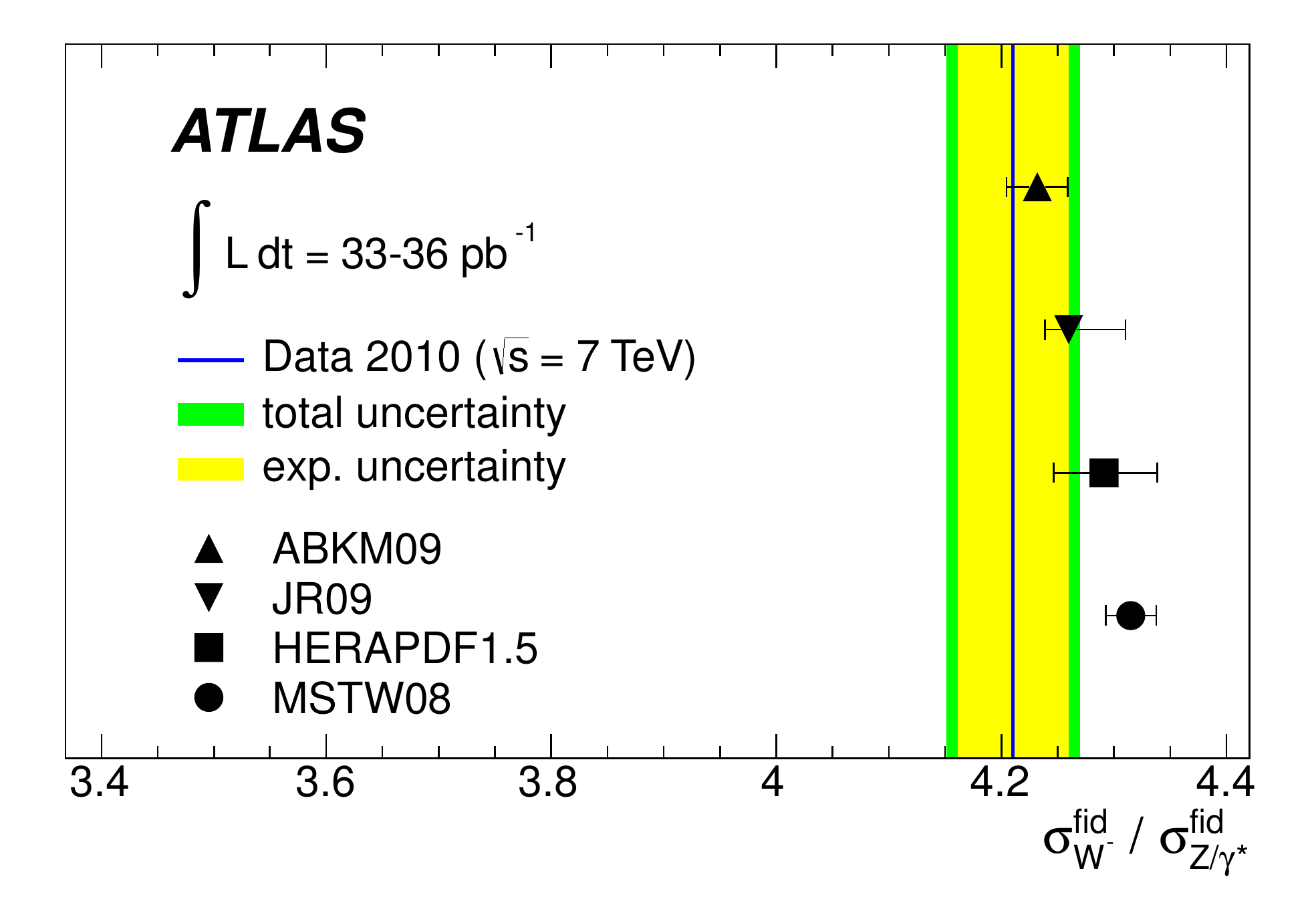}
\setcounter{figure}{2} 
\caption[]{\footnotesize   
Measured and predicted fiducial cross section ratios, $\sigma(W^+)/\sigma(Z^0)$
(left) and $\sigma(W^-)/\sigma(Z^0)$ (right). The experimental uncertainty (inner yellow 
band)
includes the experimental systematic errors. The total uncertainty (outer green band) 
includes
the statistical uncertainty and the small contribution from the acceptance correction. The
uncertainties of the ABKM \cite{Alekhin:2009ni}, JR \cite{JimenezDelgado:2008hf} and 
MSTW08
\cite{Martin:2009iq} predictions are given by the PDF uncertainties considered to 
correspond to
68 \% CL and their correlations are derived from the eigenvector sets. The results for 
HERAPDF
comprise all three sources of uncertainty of that set; from~\cite{Aad:2011dm},
\TCop (2012) by the American Physical Society.}
\end{figure}

\noindent
Due to the ratios considered the luminosity errors do widely cancel and the present
experimental accuracy nearly compares to the size of the PDF-errors. Later this year
even more precise results are expected and the analyses are still ongoing. One should also
notice, that the corresponding CMS bands \cite{CMS:2011aa} are shifted by about $1\sigma$ at 
present. For LHCb similar analyses have been performed, cf.~\cite{Aaij:2012vn}.
PDF predictions not hitting the yellow bands in Fig.~3 cannot be considered\footnote{We 
thank M. Klein for a corresponding clarification.} as outliers \cite{FW}. After all,
precision data like this will have an impact on the PDF fits very soon as has happened in
improving PDF-sets so often in the past, see e.g.~\cite{Martin:2009iq} and references therein.

Let us now consider more differential distributions in case of $W^\pm$ and $Z^0$ 
production. In Ref.~\cite{Alekhin:2013dmy} we investigated the rapidity distributions for $W^\pm, 
Z^0$-production for ATLAS and LHCb and the $e^\pm$-asymmetry for $W^\pm$ production in CMS.
This requires detailed Monte Carlo studies and comparisons based on the codes DYNNLO 1.3 
\cite{Catani:2009sm} and FEWZ 3.1 \cite{Li:2012wn}. As an example, we show the comparison of the 
ABM11 prediction for the LHCb data in Fig.~4. Tab.~7 summarizes the fit results for the
predictions based on the ABM11-distributions for ATLAS, CMS and LHCb. $\chi^2/NDP$ values of
1.3--1.07 are obtained. Here the NNLO standard value of $\alpha_s(M_Z^2)$ of ABM11 has been used. 
Contrary to this, a recent benchmarking of LHC cross sections in Table 6 of~\cite{Ball:2012wy}
reported values of $\chi^2/NDP$ in the range of 1.602-1.923 at NNLO for comparisons assuming 
$\alpha_s(M_Z^2) = 0.117$ here, which are not confirmed in \cite{Alekhin:2013dmy}. For the MSTW08 
distributions we find $\chi^2/NDP$ of 26.2/9 for the $Z$-production at LHCb. The benchmarking 
of this channel has not been performed in \cite{Ball:2012wy}.
\restylefloat{figure}
\begin{figure}[H]\centering
\includegraphics[scale=0.3,angle=0]{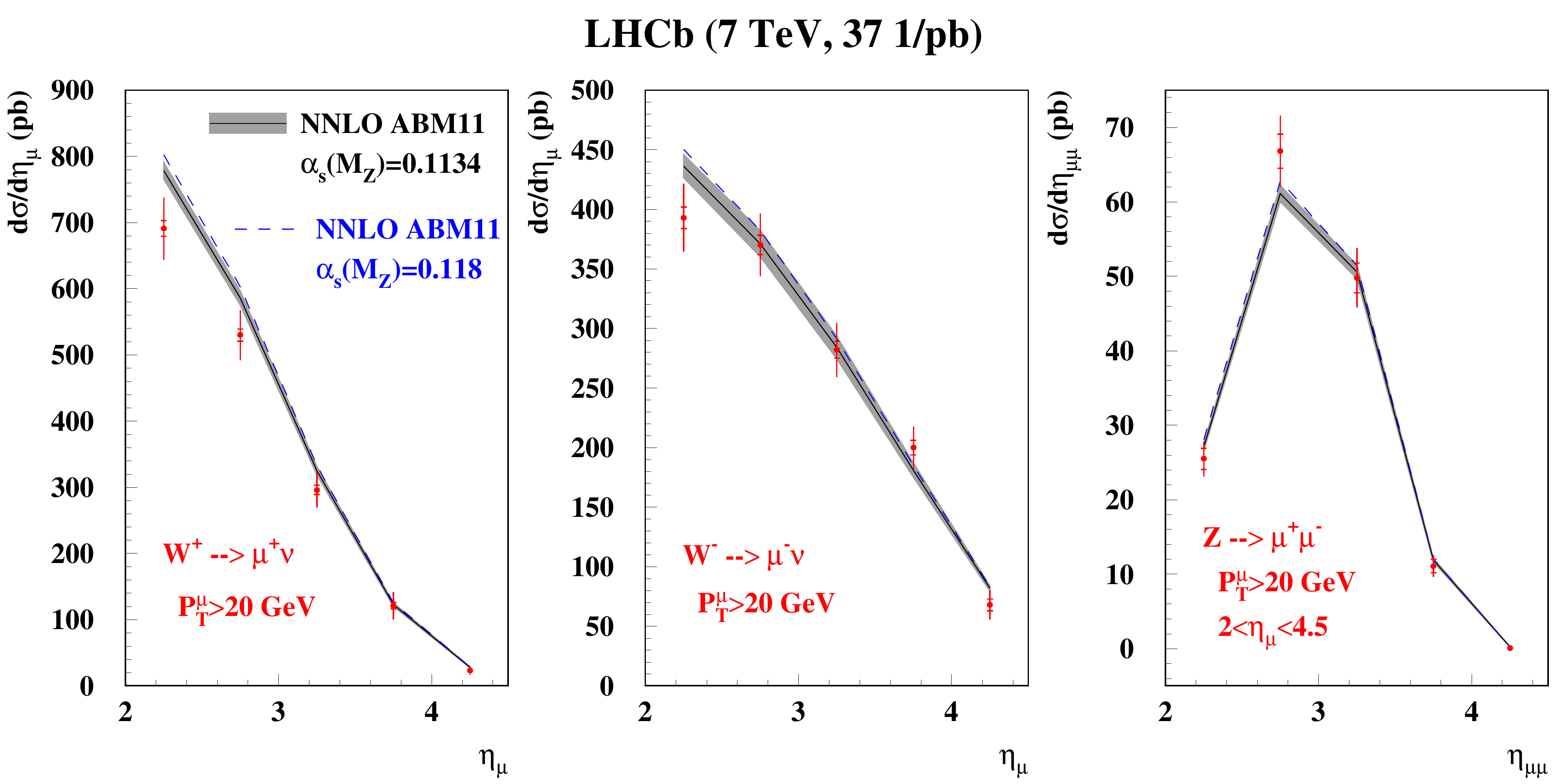}
\setcounter{figure}{3} 
\caption[]{\footnotesize  
The LHCb data of Ref.~\cite{Aaij:2012vn} on the
rapidity distribution of the charged leptons
$\mu^+$ (left panel), $\mu^-$ (central panel), and
$Z$-boson (right panel) in comparison with
the NNLO predictions computed with the ABM11 PDFs taking
into account the uncertainties due to PDFs (shaded area). The dashed curves
display the ABM11 predictions obtained with the value of
$\alpha_s(M_Z)=0.118$; from \cite{Alekhin:2013dmy}, \TCop (2013) by SISSA.}
\end{figure}

\begin{table}[th!]
\renewcommand{\arraystretch}{1.3}
\begin{center}                   
{\small                          
\begin{tabular}{|c|c|c|c|c|}   
\hline                           
{Experiment}                      
&ATLAS~\cite{Aad:2011dm}                         
&{CMS~\cite{Chatrchyan:2012xt}}  
&{LHCb~\cite{Aaij:2012vn}}
&{LHCb~\cite{lhcbe}}                            
\\                                                        
\hline
{Final states}                                                   
& $W^+\rightarrow l^+\nu$
& $W^+\rightarrow e^+\nu$
&$W^+\rightarrow \mu^+\nu$
& $Z\rightarrow e^+e^-$                                                        
\\                                                        
& $W^-\rightarrow l^-\nu$
&$W^-\rightarrow e^-\nu$
&$W^-\rightarrow \mu^-\nu$
&                                                         
\\                                                        
& $Z\rightarrow l^+l^-$
&                                                         
&                                                         
&                                                         
\\                                                        
\hline                                                    
{Luminosity (1/pb)}                      
&35                         
&840  
&{37}
&{940}                            
\\                                                        
\hline                                                    
$NDP$
&30                      
&11  
&10
&9                            
\\                                                        
\hline
 $\chi^2$
 &$34.7(7.7)$
 &$11.8(4.7)$
 &13.0(4.5)
 &11.5(4.2)
\\
\hline                                          
\end{tabular}
}
\setcounter{table}{6} 
\caption{\footnotesize The value of $\chi^2$ obtained for different samples of 
the Drell-Yan LHC data with the NNLO ABM11 PDFs.
The figures in parenthesis give one standard deviation of 
$\chi^2$ equal to $\sqrt{2NDP}$; from \cite{Alekhin:2013dmy} \TCop (2013) SISSA.}
\end{center}
\label{tab:chi2}
\end{table}

At present the inclusive $t\bar{t}$-production cross section is measured by 
ATLAS and CMS with
\begin{eqnarray}
\label{sigtt1}
\sigma(t\bar{t}) &=& 186 \pm 13~{\rm (stat.)} \pm 20~{\rm (syst.)} \pm  7~{\rm (lum.)}~  
pb~~{\rm ATLAS},~~[63] 
\\
\sigma(t\bar{t}) &=& 161.9 \pm  2.5~{\rm (stat.)} \pm~5.1~{\rm (syst.)} \pm 3.6~{\rm (lum.)}~ 
pb~~{\rm 
CMS}~~[64] 
\\
\label{sigtt2}
\sigma(t\bar{t}) &=& 139 \pm  10~{\rm (stat.)} \pm  26~{\rm (syst.)} \pm 3~{\rm (lum.)}~pb~~{\rm 
CMS, jets}~~[65], 
\end{eqnarray}
for $m_t = 172.5$~GeV.
The current difference between both experiments has still 
to be understood and reconciled. Furthermore, the calculation of the inclusive 
$t\bar{t}$--production cross section
at NNLO has to be completed. The scattering cross section exhibits a very strong dependence 
on the top quark mass $m_t$. In the ABM11 analysis a value of $\sigma(t\bar{t}) =
130.4 {\small \begin{array}{c} + 2.9 \\ -7.2 \end{array}}~{\rm (scale)} \pm 5.9~{\rm 
(PDF)}$~\footnote{
In \cite{FW} the cross section for ABM11 \cite{Alekhin:2012ig} is quoted by 7 $pb$ too low.}  
is obtained for $m_t = 173$ GeV
using {\tt HATHOR}~\cite{Aliev:2010zk},
with a variation form $\sigma(t\bar{t}) = 167.9 ... 122.6~pb$ for $m_t$ in the range 
[$165,175$~GeV].
On the other hand, JR, MSTW08 and NN21 give values in the range of $159.3-167.7~pb$ for $m_t = 
173$~GeV. In this comparison, the  ABM11 value comes out lower.
With a variation to somewhat lower values of $m_t$ it is also getting 
closer to the higher values of CMS and ATLAS and consistent with the lower value by CMS. In the 
future a final LHC analysis will deliver a unique value for 
$\sigma(t\bar{t})$ with smaller errors and a determination of $m_t^{\overline{\rm MS}}$ being fully 
consistent with the quantum field-theoretic description of the scattering cross section. 

The inclusive Higgs-boson production cross section is dominated by the gluonic channel and depends
on both the gluon distribution function and the strong coupling constant as 
$\propto \alpha_s^2(M_H^2) xG(x,M_H^2) \otimes xG(x,M_H^2)$, with $M_H$ the Higgs boson mass and 
$\otimes$ denoting the Mellin convolution. Due to this the errors both in $\alpha_s$ and $xG$ enter 
the cross section twice and a precise prediction requires that both quantities have to be 
understood very accurately. Detailed tables of the production cross sections have been given in
Refs.~\cite{Dittmaier:2011ti,Alekhin:2010dd,Dittmaier:2012vm,Alekhin:2012ig} and in 
Ref.~\cite{Anastasiou:2012hx}. After first evidence for a Higgs-like particle has been found 
in the mass range $\sim 125$ GeV \cite{Aad:2012tfa,Chatrchyan:2012ufa}, the
search is ongoing and further detailed results are expected to be presented for the $\sqrt{s} = 7$ 
and 8 TeV runs at LHC very soon. 
\section{Conclusions}

\noindent
We reported on recent results of NNLO determinations of the parton distribution functions
and of $\alpha_s(M_Z^2)$, based on the world precision data on deep-inelastic scattering
and sensitive data from hadron colliders as Tevatron and the LHC, comparing the results
obtained by different fitting groups. There are still differences being observed. In case
of the gluon distribution the upcoming precision measurement of the jet cross sections
at LHC may deliver a new impact. In various analyses low values of $\alpha_s(M_Z^2)$ are obtained in 
NNLO,
for which we quote the value of ABM 11 with $\alpha_s(M_Z^2) =  0.1134  \pm 0.0011$.
We have given reasons why the NNLO values for $\alpha_s(M_Z^2)$ are coming out larger
for MSTW08 and NNPDF and think that the current difference is understood. The correct inclusion of 
systematic 
error correlations and higher twist effects are important. The inclusion of the ATLAS jet 
data into the present analysis leads to $\alpha_s(M_Z^2) = 0.1141 \pm 0.0008$. 
The gluon distribution being obtained is softer than that in case of fitting to  the Tevatron 
jet data. 
A rather consistent picture has been obtained for the inclusive $W^\pm$ and $Z^0$ production
cross section ratios by ATLAS, CMS and LHCb, which show some sensitivity to the quark and 
anti-quark densities. For the inclusive $t\bar{t}$-cross section there are still differences 
between the measurements at ATLAS and CMS. The different fitting groups also report different
results here. However, the scattering cross sections do strongly depend on $\alpha_s(M_Z^2), 
xG(x,Q^2)$ and the value of $m_t$ assumed in the data analysis. Furthermore, the NNLO corrections 
are not yet complete for this process. 

The gluon density will soon be constrained by the present LHC data
on two and three jet production. These data will also yield an improved value of
$\alpha_s(M_Z^2)$. There is a need of continuous benchmarking between groups 
using correct statistical methods. An effective rescaling of experimental errors, 
even with individual factors changing for different data sets, is neither necessary
nor can it lead to an objective result. It is evident that highly precise and
systematically well-controlled data sets have to be given preference over data
with a poorer systematics in global analyses to determine the parton distribution
functions and $\alpha_s(M_Z^2)$. 

The so-called `pdf4lhc recommendation' \cite{Botje:2011sn} unfortunately left out two out 
of three NNLO 
analyses since 2009. Moreover, the effect of the combined HERA data has not been considered 
in the fits chosen there. On the other hand, all NNLO analyses are available in terms of 
parameterizations in LHAPDF and most of them run very fast, which is important for 
experimental use. We therefore recommend that the experimental groups {\it freely} use all 
NNLO PDFs within their analyses together with the correlation matrices and the right value of 
$\alpha_s(M_Z^2)$. Since more and more differential distributions are investigated with a need
of sophisticated systematic analyses by the LHC experiments, external determinations of theory 
errors related to the available PDFs become readily impossible. They have to be determined
by the experiments to deliver unbiased precision results. The benchmarking between the different 
fitting groups has to continue to finally provide a thoroughly agreeing picture both on the PDFs
and on $\alpha_s(M_Z^2)$ from DIS and precise hard scattering cross sections at high energies.

\vspace*{1mm}
\noindent
{\bf Acknowledgment.}~We would like to thank N. Glover, M. and U. Klein, J. Mnich, 
K.~M\"uller, K. Rabbertz and E. Reya for discussions and J. Mnich for reading the manuscript. 
This work has been supported in part by DFG Sonderforschungsbereich Transregio 9, Computergest\"utzte 
Theoretische Teilchenphysik, Helmholtz Alliance at the Terascale, and by the EU Network 
{\sf LHCPHENOnet} PITN-GA-2010-264564.

\vspace*{1mm}
\noindent
{\sf Note added.}~~After completion of the present paper
a preprint with details on the CT10 analysis appeared \cite{CT10a}. Here 
$\alpha_s(M_Z)$ is dealt with as a pilot parameter at NNLO and not fitted. 
As known from other analyses, for a choice of $\alpha_s(M_Z) \in [0.112,0.127]$ 
very different values of $\chi^2_{\rm min}$ are obtained, cf. e.g. Figs.~4.16--4.18 of 
Ref.~\cite{Alekhin:2012ig}.
 

\fi
\end{document}